\documentclass[aps,prb,reprint,longbibliography]{revtex4-2}
\usepackage{graphicx}
\usepackage{amsmath}
\usepackage{amssymb}
\usepackage{natbib}
\usepackage{hyperref}
\usepackage{float}
\usepackage{xcolor}
\usepackage{ulem}
\begin{document}
\title{Magnetic and ferroelectric phase diagram of twisted CrI$_3$ layers}
\author{Haoshen Ye}
\author{Shuai Dong}
\email{sdong@seu.edu.cn}
\affiliation{Key Laboratory of Quantum Materials and Devices of Ministry of Education, School of Physics, Southeast University, Nanjing 21189, China}
\date{\today}

\begin{abstract}
Twisting layers provide a rich ore for exotic physics in low dimensions. Despite the abundant discoveries of twistronics from the aspect of electronic structures, ferroic moir\'e textures are more plain and thus less concerned. Rigid lattice models are straightforward which can give a rough but intuitional description in most cases. However, taking CrI$_3$ as a model system, here we will demonstrate that the interlayer stacking potential can spontaneously lead to structural relaxation, which plays a vital role to understand the ferroicity in the twisted superlattices. The magnetic ground state is sensitive to the stacking mode and twisting angles, which can be seriously affected by the structural relaxation. In particular, the expected magnetic bubbles are annihilated in its bilayer. In contrast, because of topological protection, the ferroelectric vortices are more robust to structural relaxation, as well as twisting angle and thickness. Because of the universal existence of spontaneous structural relaxation in twisted superlattices, our work may lead to a general revisitation of emerging physics of twistronics.
\end{abstract}
\maketitle

Magnets and ferroelectrics are key functional materials widely used in technological applications. The rise of van der Waals (vdW) materials opens new avenues for realizing magnetism and ferroelectricity in reduced dimensions \cite{N-2017-CrI3, N-2017-CrGeTe, NC-2016-CIPS, N-2022-NiI2, CR-2023-review}. The weak interlayer interactions make these vdW multilayers highly tunable, which allows for the manipulations of their electronic, magnetic, and ferroelectric properties through external stimuli \cite{NM-Yankowitz-2014-Gr, S-Song-2018-CrI3, S-Ji-2022-TMDs, NM-Song-2019-CrI3, PRL-Wang-2024-BN}. Such tunability paves the way for novel applications with versatile quantum functionalities \cite{NP-Xiao-2020-WTe2, PRL-Yang-2024-BN, PRL-Yang-2023-BN, PRL-2024-AM}.

Twist engineering, by creating angle mismatch between adjacent vdW layers, has been found to be a rich ore of exotic physics, such as ultraflat bands, topological polaritons, as well as unconventional superconductivity \cite{NC-2024-flatband, N-2020-MoO3, N-2018-Gr}. Besides these holistic effects, the periodic variation of stacking modes will induce spatial modulation of magnetism, ferroelectricity, as well as optical response in twisted superlattices \cite{S-Song-2021-CrI3, N-2024-BTO,S-Yasuda-2021-BN, NP-2021-TMDs, NC-2024-BP}. In fact, a twisted superlattice can be understood as a mosaic of local stacking modes \cite{CR-2023-review, PRB-2018-config-space, npj2D-2022-config-space, NL-2021-CrX3}. Because of the one-to-one correspondence between the stacking mode and some physical properties, one can naturally expect the formation of periodic ferroic domains in these twisted superlattices, providing an ideal platform to study and manipulate the ferroic functionalities of domains and domain walls \cite{NC-2023-BN, PNAS-2020-phases}.

Another consequence of twisting is that the local stacking energy is also periodically modulated. To reduce the stacking energy, structural relaxation will be inevitable, which has been experimentally evidenced in twisted superlattices \cite{NC-Li-2021-WSe2, NN-2020-TMDs, NL-2023-MoS2, NM-Yoo-2019-Gr}. However, its effects to magnetism and ferroelectricity in twisted vdW superlattices have been mostly ignored, while simplified rigid lattices are usually modeled to describe these twisted layers \cite{NL-2021-CrX3, NC-2023-BN, PNAS-2020-phases}. Then an essential question is that whether the structural relaxation is vital to the ferroic physics in twisting layers, or it is indeed negligible as done before. This issue should be solved before the next step of twistronics with ferroicity involved.

In this Letter, we will clarify the key role of structural relaxation to magnetism and ferroelectricity in twisted CrI$_3$ superlattices, going beyond the rigid model. On one hand, the magnetic texture of twisted CrI$_3$ layers is found to be significantly affected by the structural relaxation, which can fully suppress the magnetic bubbles in the CrI$_3$ bilayer with small twisting angles. On the other hand, although the ferroelectric domain walls are also seriously squeezed, the ferroelectric vortices survive after the structural relaxation, due to the topological\sout{ly} protection. These results solve the discrepancy between prior experimental observations and theoretical expectations \cite{NN-2022-t2LCrI3, PNAS-2020-phases, NC-2023-BN}, providing a complete description of ferroicity in twisted materials.

As one of the earliest two-dimensional (2D) ferromagnets confirmed in experiment \cite{N-2017-CrI3}, CrI$_3$ has received numerous attentions and become the most studied 2D magnet. Interestingly, the sign of its interlayer magnetic coupling depends on the stacking mode \cite{NM-Song-2019-CrI3}, and sliding ferroelectricity is also expectable by tuning the stacking modes \cite{PRL-Ji-2023-CrI3}. These stacking-dependent ferroic properties make twisted CrI$_3$ a perfect model system for studying the effects of structural relaxation to ferroicity in twisted superlattices.

\begin{figure}
\centering
\includegraphics[width=0.48\textwidth]{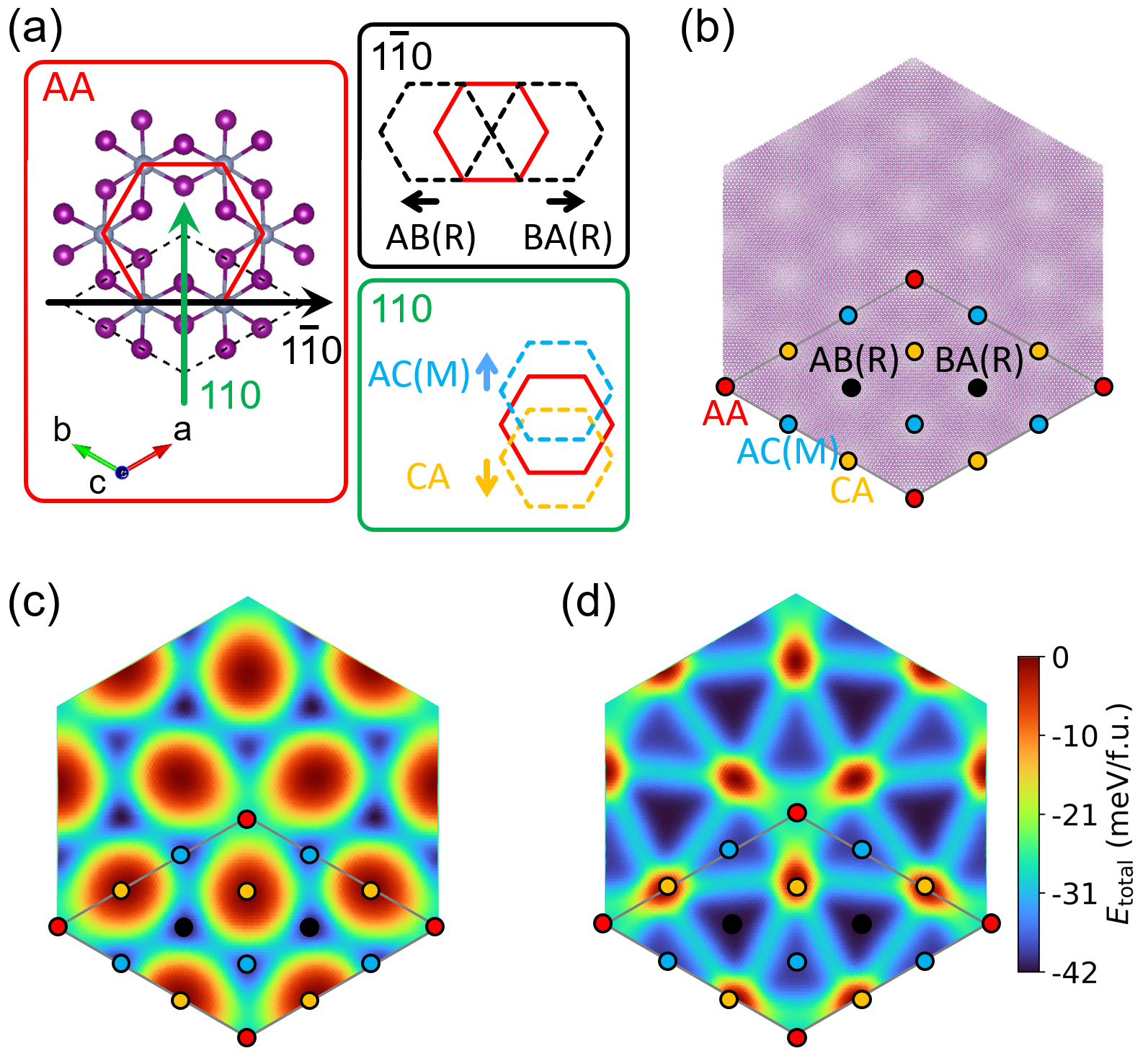}
\caption{The $1+1$ CrI$_3$ bilayer. (a) Typical stacking modes obtained by interlayer sliding. (b) Atomic structure of the $\theta_{\rm twist}=0.99^\circ$ rigidly twisted case. Rhomb with broken lines: the minimum supercell. Color dots: the special positions corresponding to particular stacking modes. (c-d) Total energy ($E_{\rm total}$) landscapes of the $\theta_{\rm twist}=0.99^\circ$ case: (c) without and (d) with structural relaxation.}
\label{Fig1}
\end{figure}

Let us start from the simplest case: the CrI$_3$ $1+1$ bilayer. As illustrated in Fig.~\ref{Fig1}(a), the AA stacking mode can be transformed to the AB (or BA) mode by laterally sliding the top layer along ($1\overline{1}0$) by $-1/3$ (or $1/3$) in fractional coordinate. By sliding along ($110$) by $-1/3$ and $1/3$, one can obtain the AC and CA stacking modes, respectively. In fact, the AB/BA and AC modes correspond to the low-temperature rhombohedral ($R$) and high-temperature monoclinic ($M$) phases of bulk CrI$_3$ \cite{CM-2015-CrI3}. The stacking energy landscape of sliding operation can be obtained via density functional theory (DFT) calculation (see EM1 in End Matter), which is consistent with previous studies \cite{NL-Sivadas-2018-CrI3, PRL-Ji-2023-CrI3}. The details of DFT calculations can be found in Supplemental Material \cite{sm}.

As aforementioned, the twisting operation, characterized by the twisting angle $\theta_{\rm twist}$, can tune the stacking degree of freedom. For example, a $\theta_{\rm twist}=0.99^\circ$ twisting of CrI$_3$ bilayer leads to a supercell containing $13~468$ formula units, as Fig.~\ref{Fig1}(b). The typical stacking modes can also be evidenced locally in the twisted bilayer. The CrI$_3$ monolayer belongs to the space group $P\overline{3}1m$, which contains the threefold rotation ($C_{3z}$) and vertical mirror ($\sigma_v$) symmetries. The $C_{3z}$ symmetry ensures that every $120^\circ$ rotation is equivalent to $0^\circ$. The $\sigma_v$ symmetry makes the mirror-symmetric atomic positions of clockwise and anti-clockwise rotation by same angles relative to $0^\circ$. Consequently, the study of $\theta_{\rm twist}\in[0^\circ$, $60^\circ]$ captures all physics for twisted CrI$_3$ bilayer.

For the twisting mode shown in Fig.~\ref{Fig1}(b), the distribution of stacking energy $E_{\rm stack}$ can be mapped to Fig.~\ref{Fig1}(c). The gradient of stacking energy leads to intrinsic strain and stress, which will lead to spontaneous structural relaxation till the balance between $E_{\rm stack}$ and the strain energy $E_{\rm strain}$. $E_{\rm strain}$ can be expressed as:
\begin{equation}
E_{\rm strain} = \frac{1}{2}\sum_{ij} C_{ij}\varepsilon_{ij}^2,
\label{E_e}
\end{equation}
where $C$ and $\varepsilon$ are the elastic tensor and strain tensor, respectively. $i$/$j$ are coordinate axes. The value of $C$ can be obtained via DFT calculation (Fig. S2 (a) and Table S1 in Supplemental Material) \cite{sm}. 

By minimizing the total energy $E_{\rm total}=E_{\rm strain}+E_{\rm stack}$, the relaxation can be numerically simulated (see EM1 for details). As shown in Fig.~\ref{Fig1}(d), the CA regions with high $E_{\rm stack}$ profile seriously shrink to accommodate more $R$-phase and $M$-phase domains which are more energetically favorable. For the $\theta_{\rm twist}=0.99^\circ$ case, the total energy of CrI$_3$ bilayer is reduced by $9.85$ meV/f.u. ($\Delta E_{\rm stack}$: $-9.94$ meV/f.u. and $\Delta E_{\rm strain}$: $+0.09$ meV/f.u.) from the structural relaxation, which is only $11.87$ meV/f.u. higher than the lowest-energy AB/BA stackings. The increased $E_{\rm strain}$ comes from the strain accumulation at domain walls (see Figs.~S3 and S4 in Supplemental Material \cite{sm}). This relaxation is weakened in larger $\theta_{\rm twist}$ cases with smaller supercells and domains. While in smaller $\theta_{\rm twist}$ cases with larger supercells and domains, the width of stacking domain walls eventually stabilizes at an optimal value ($\sim97\pm1$ \AA).

\begin{figure}
\centering
\includegraphics[width=0.48\textwidth]{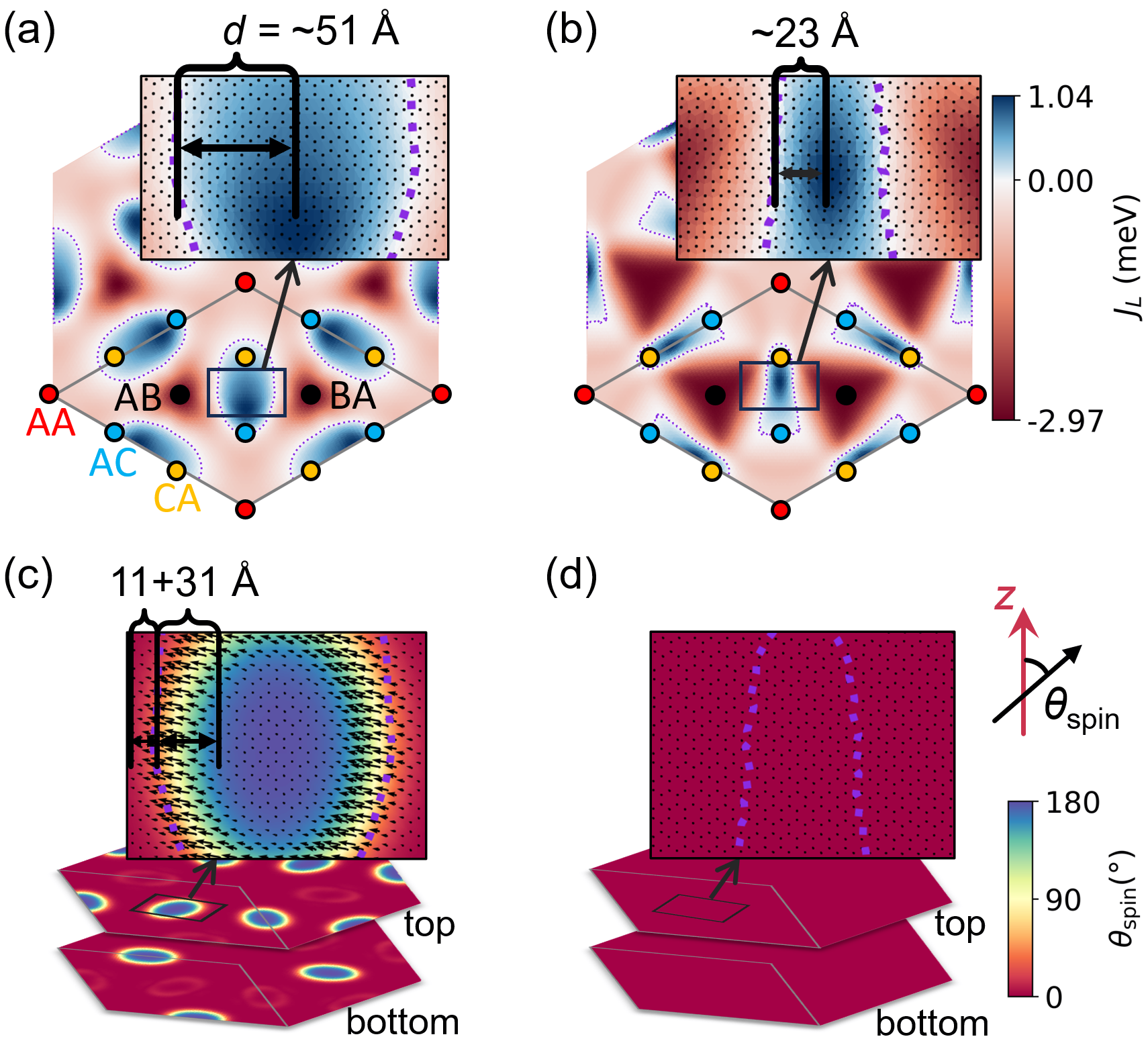}
\caption{Contour maps of interlayer magnetic coupling of $\theta_{\rm twist}=0.99^\circ$ $1+1$ CrI$_3$ bilayers: (a) without and (b) with structural relaxation. Negative and positive values of $J_L$ denote ferromagnetic and antiferromagnetic interactions. Corresponding magnetic textures: (c) without and (d) with structural relaxation. $\theta_{\rm spin}$: the polar angle of spin as the inset illustrates. Purple broken curve: the $J_L=0$ boundary.}
\label{Fig2}
\end{figure}

For CrI$_3$, the sign of interlayer magnetic coupling ($J_L$) is sensitive to the stacking mode \cite{NL-Sivadas-2018-CrI3, PRL-Ji-2023-CrI3}. As shown in Fig.~\ref{Fig2}(a), the elliptic regions with AC/CA-like stacking favor the antiferromagnetic coupling, while the AB/BA-like regions strongly favor the ferromagnetic coupling. The spatial modulation of $J_L$'s sign in this twisted bilayer creates magnetic competition around the transition boundary, which have been reported as a promising platform for magnetic bubbles (and merons) \cite{NL-2023-CrCl3, NCS-Yang-2023-CrI3}. Comparing with the pristine case, the antiferromagnetically-preferred regions are seriously shrunk [Fig.~\ref{Fig2}(b)] after the structural relaxation, which may affect the stability of domain walls. In fact, sufficient space is required to accommodate the magnetic domain walls, otherwise, magnetic bubbles cannot survive.

To estimate the width of magnetic domain walls and analyze the spin textures in the twisted CrI$_3$ bilayer, micromagnetic simulations based on the Landau-Lifshitz-Gilbert (LLG) equation is performed (see EM2 for details). As shown in Fig.~\ref{Fig2}(c), an elliptic magnetic bubble is formed at each antiferromagnetic $J_L$ region, where spins are $180^\circ$ flipped cross the magnetic domain wall with a width of $\sim42$ {\AA} in the rigidly twisted CrI$_3$ bilayer. By deducting the distance from the $J_L=0$ boundary to the domain wall boundary ($\sim11$ {\AA}), the required half-width ($d$) of the antiferromagnetic $J_L$ region should be wider than $31$ {\AA} to meet the space requirement of forming a magnetic bubble in the rigidly twisted CrI$_3$. As expected, for the relaxed structure, the $J_L>0$ region is too narrow to host any magnetic bubble, as shown Fig.~\ref{Fig2}(d).

\begin{figure}
\centering
\includegraphics[width=0.48\textwidth]{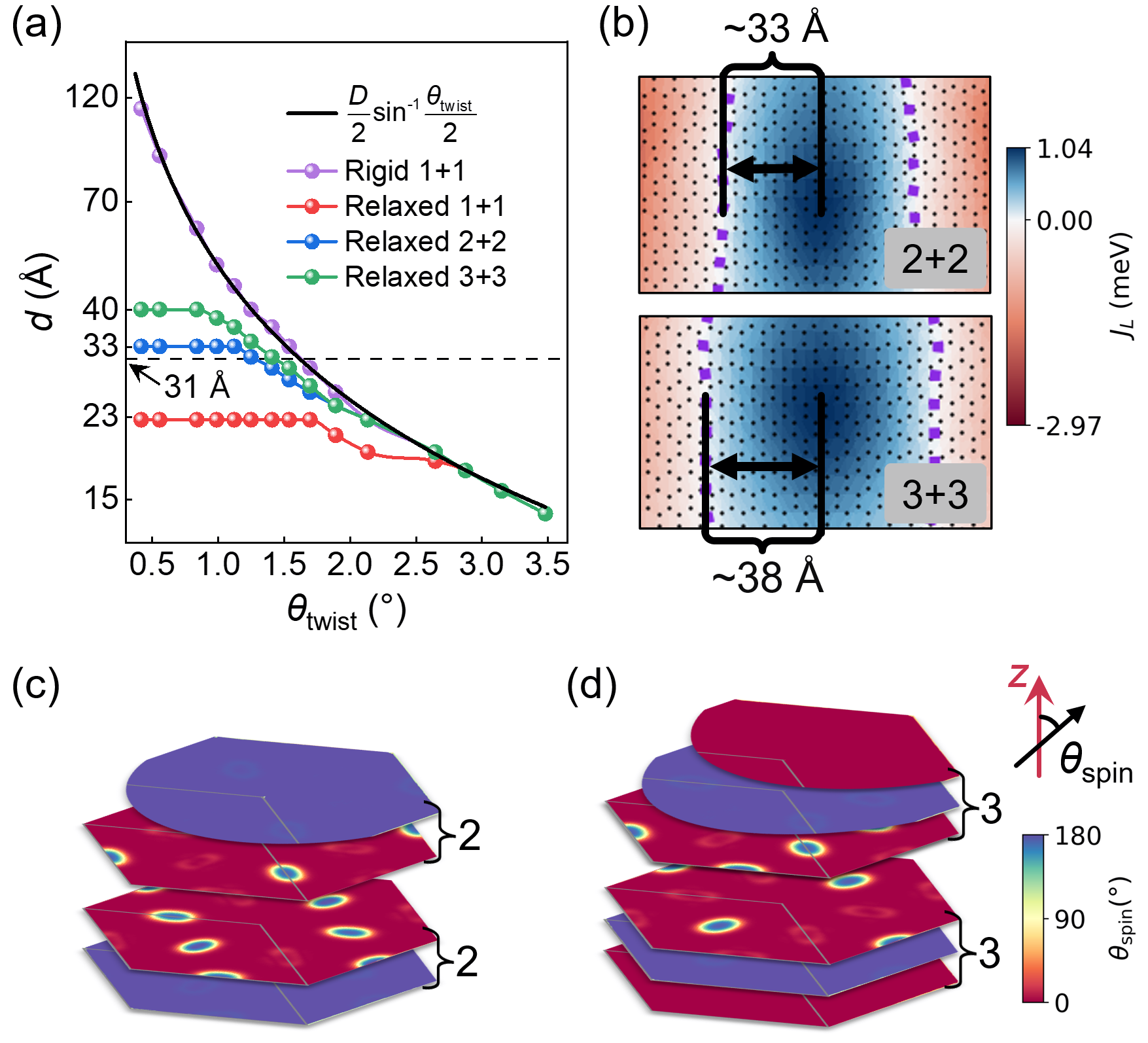}
\caption{Tuning of structural relaxation in twisted CrI$_3$ layers. (a) Evolution of the half-width $d$ of antiferromagnetic $J_L$ region as a function of $\theta_{\rm twist}$ for twisted CrI$_3$ bilayers. Broken line: the threshold to accommodate a magnetic bubble. (b-d)  The $\theta_{\rm twist}=0.99^\circ$} $2+2$ and $3+3$ CrI$_3$ layers after structural relaxation. (b) Contour maps of $J_L$. Their $d$'s are wide enough for magnetic bubbles. (c-d) The corresponding spin textures obtained from simulation.
\label{Fig3}
\end{figure}

Then a following question is that whether any magnetic bubble can survive by tuning the twisting, e.g. $\theta_{\rm twist}$ and layer thickness. To clarify this issue, we start from the rigidly twisted model. According to the direct geometric triangulation, the half-width $d_{\rm rigid}$ of antiferromagnetic $J_L$ region in the rigidly twisted  $1+1$ CrI$_3$ bilayer can be estimated as:

\begin{equation}
d_{\rm rigid}=\frac{D}{2}\sin^{-1}(\frac{\theta_{\rm twist}}{2}), 
\label{d}
\end{equation}
where the scaling factor $D=0.88$ {\AA} (see EM3 for details). As shown in Fig.~\ref{Fig3}(a), the analytic Eq.~(\ref{d}) indeed matches the numerical solution of unrelaxed $1+1$ CrI$_3$ bilayer. $\theta_{\rm twist}$ should be less than $1.54^\circ$ to provide enough $d$ ($>31$ \AA) to accommodate magnetic bubbles in the rigidly twisted $1+1$ CrI$_3$ bilayer. This estimation is consistent well with our numerical simulations (see Fig.~S6 \cite{sm}). 

After the structural relaxation, the real $d$ is always smaller than $31$ \AA{} for any $\theta_{\rm twist}$, which reaches a saturation value ($\sim23$ \AA) in the small $\theta_{\rm twist}$ limit. This result solves the puzzle that why there is no magnetic bubbles observed in the twisted $1+1$ CrI$_3$ bilayer~\cite{NN-2022-t2LCrI3}, and predicts that it will never be while it should appear in the small $\theta_{\rm twist}$'s cases according to the rigid model.

In the large $\theta_{\rm twist}$ limit, the supercell becomes smaller and smaller, which leads to smaller modulation periods and thus smaller $d$. Then the structural relaxation will be gradually suppressed, since the energy gain of $E_{\rm stack}$ from relaxation becomes meager to compensate the rapid increasing of $E_{\rm strain}$ (Fig.~S2 in Supplemental Material \cite{sm}). Indeed, when $\theta_{\rm twist}>2.65^\circ$, the simulated $d$ is close to $d_{\rm rigid}$. 

Above study has demonstrated that the structural relaxation is disadvantage for the appearance of magnetic bubble. A straightforward way to suppress the structural relaxation is to tune the balance between $E_{\rm stack}$ and $E_{\rm strain}$. A simple way is to use thicker twisting layers, in which the effective elastic stiffness will be significantly enhanced. Then the structural relaxation will become weaker.

To verify this hypothesis, we investigate the twisted $L+L$ bilayers ($L=2$ and $3$), by keeping the top and bottom $L$-layer in the $M$ stacking mode as observed in experiment \cite{NE-2023-tDCrI3, NC-2024-3LCrI3}. In these configurations, the interfacial stacking energy $E_{\rm stack}$ remains unchanged, while the strain energy $E_{\rm strain}$ is doubled or tripled compared to that of $L=1$ case. As a result, the structural relaxation is naturally suppressed in these thick cases, making the relaxed structures and the $J_L$ distributions closer to the rigid lattice model. Therefore, $d>31$ {\AA} remains a valid criterion to host magnetic bubbles (see Figs.~S7-S9 in Supplemental Material \cite{sm}). As compared in Fig.~\ref{Fig3}(a), larger $d$'s are reserved for the $L=2$ and $3$ cases, which can go beyond the threshold in the small $\theta_{\rm twist}$ limit ($1.25^\circ$ for $L=2$ and $1.34^\circ$ for $L=3$). The simulation patterns of $J_L$'s distribution confirm this scenario, as shown in Fig.~\ref{Fig3}(b). As expected, $d$ reaches $\sim33$ {\AA} ($L=2$) and $\sim38$ {\AA}  ($L=3$) for the $\theta_{\rm twist}=0.99^\circ$ cases. Consequently, the magnetic bubbles indeed appear in these thick twisted bilayers, as shown in Figs.~\ref{Fig3}(c-d), consistent with the experimental observation of twisted double-bilayer and double-trilayer of CrI$_3$ \cite{S-Song-2021-CrI3, NC-2024-CrI3,NC-2024-3LCrI3}.

Another interesting observation is that the magnetic bubbles only appear in two interfacial layers, instead of magnetic columns running through the bilayer or trilayer. The number of these 2D bubbles is a constant: three per supercell. Due to the reciprocal relationship between $\theta_{\rm twist}$ and the size of supercell, those $\theta_{\rm twist}$'s close to the threshold limit can lead to relative high density of magnetic bubbles.

\begin{figure}
\centering
\includegraphics[width=0.48\textwidth]{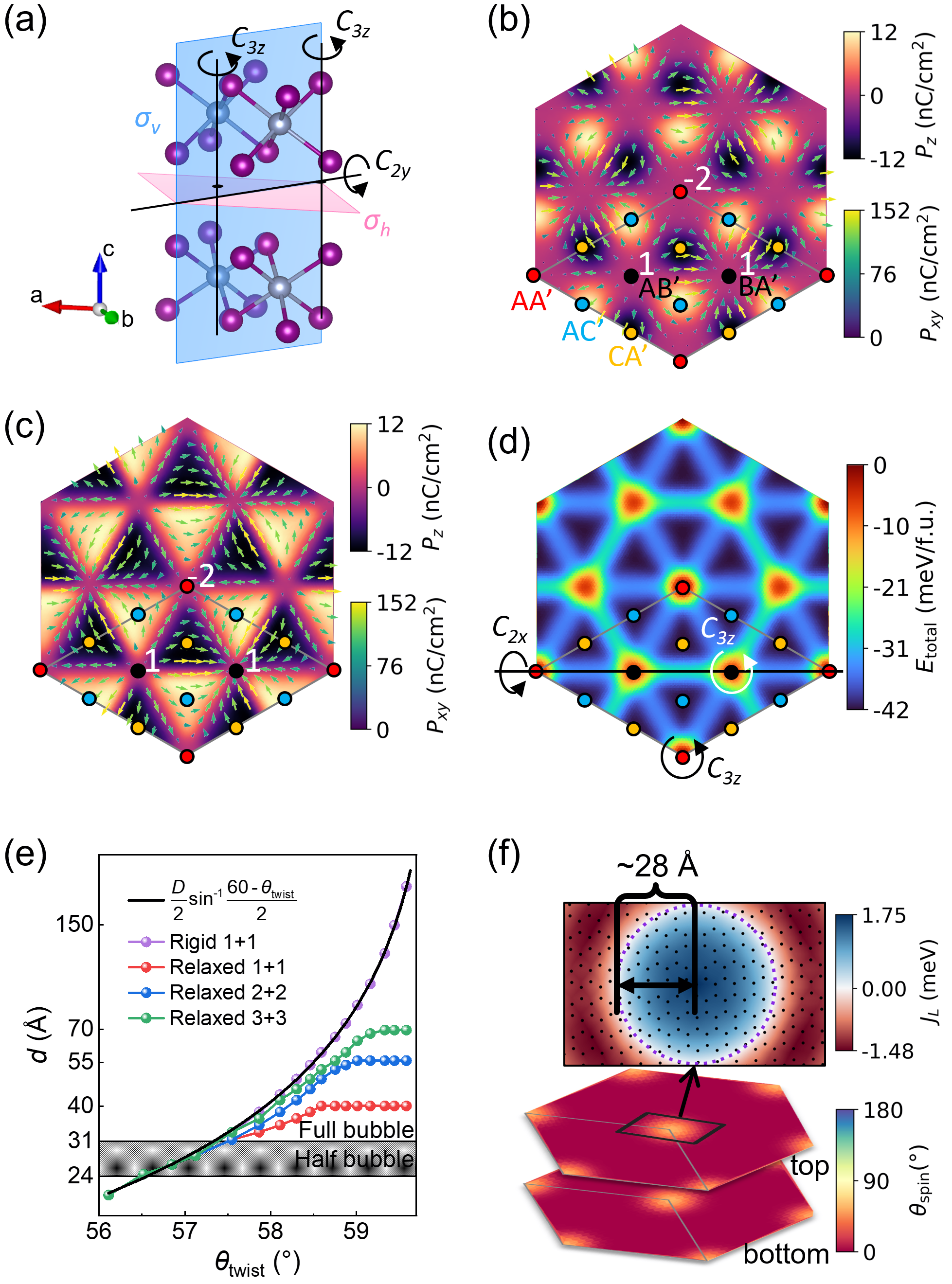}
\caption{The $\theta_{\rm twist}\sim60^\circ$ twisted (AA'-like) $1+1$ CrI$_3$ bilayer. (a) Symmetric elements of the AA' stacking.  Ferroelectric textures of the $\theta_{\rm twist}=59.01^\circ$ case: (b) without and (c) with structural relaxation. Color dots: the high-symmetric stacking points. Color arrows: the direction of local in-plane electric polarization ($P_{xy}$), while the out-of-plane component ($P_z$) is shown as the contour map. The winding numbers ($1$ \& $-2$) of ferroelectric vortices are indicated. (d) The energy landscape of the $\theta_{\rm twist}=59.01^\circ$ case and the symmetric elements. (e) Evolution of $d$ as a function of $\theta_{\rm twist}$ for twisted CrI$_3$. The thicker $2+2$ and $3+3$ cases are shown for comparison. (f) Magnetic texture and $J_{L}$ of the $\theta_{\rm twist}=56.85^\circ$ case.}
\label{Fig4}
\end{figure}

Above study has demonstrated that nontrivial magnetic bubbles as the ground state may survive in the small $\theta_{\rm twist}$ limit, while the ground state turns to be plain ferromagnetic when $\theta_{\rm twist}$ is slightly large. However, when $\theta_{\rm twist}$ is too large, e.g. close to $60^\circ$, there is another story of ferroicity. In fact, the $\theta_{\rm twist}=60^\circ$ equals to the AA' stacking (i.e. the reversed stacking), which cannot be obtained via pure sliding operations from the AA one. In the following, we will study the case with $\theta_{\rm twist}=(60-\delta)^\circ$, where $\delta$ is a small quantity. 

The most interesting physics of reversed stacking is the sliding ferroelectricity, which is absent in the AA one and its derivatives via sliding operations. The $1+1$ AA' stacking CrI$_3$ bilayer adopts the noncentrosymmetric nonpolar space group $P\overline{6}m2$, and the AB' and BA' ones own the nonpolar space group $P321$ which are obtained by the interlayer sliding along ($1\overline{1}0$) by $\pm1/3$ in fractional coordinate. As shown in Fig.~\ref{Fig4}(a), there are a horizontal mirror ($\sigma_h$) plane, three vertical mirror ($\sigma_v$) plane, three $C_{2y}$ axes, and two $C_{3z}$ axes in the AA' stacking CrI$_3$ bilayer. For all other stackings in the AA'$\rightarrow$AC'$\rightarrow$CA'$\rightarrow$AA' path, both $P_{xy}$ and $P_z$ are allowed within the $\sigma_v$ mirror plane. For sliding perpendicular to the $\sigma_v$ plane, i.e. along the AA'$\rightarrow$AB'$\rightarrow$BA'$\rightarrow$AA' path, only the $C_{2y}$ axis is preserved. Thus the allowed $P_{xy}$ is aligned along the $C_{2y}$ axis and $P_{z}=0$. For any other sliding, the $C_{2y}$ symmetry is broken and both $P_z$/$P_{xy}$ components are allowed. The distributions of $P_{z}$ and $P_{xy}$ are summarized in Fig.~\ref{Fig4}(b). The largest $P_z=12$ nC/cm$^2$ is obtained in the AC'/CA' stackings, while the largest $P_{xy}=152$ nC/cm$^2$ is achieved in the intermediate of AC'$\rightarrow$CA' path. 

The nonpolar AA'/AB'/BA' stackings are energetically unfavorable [Fig.~\ref{Fig6}(b) in EM1], which will lead to structural relaxation. As shown Fig.~\ref{Fig4}(c), after relaxation, the dominant stacking domains are alternating AC'/CA' ones, and the residual AA'/AB'/BA' modes serve as nodes connected by the AC'/CA' stacking domain walls. Considering the texture of $P_{xy}$ [Figs.~\ref{Fig4}(b-c)], there are two types of AC'/CA' stacking domain walls: a higher energy one connecting the nearest AB' and BA' nodes; a lower energy one connecting the AA' and AB'/BA' nodes. The ratio of these two domain walls are $1:2$, as depicted in Fig.~\ref{Fig4}(d). As the symmetric requirement, all domain walls align parallelly to the $C_{2x}$ axes. Therefore, the polarizations of these stacking domain walls should be along the domain walls with $P_z=0$. 

Around the nonpolar AA'/AB'/BA' nodes, ferroelectric vortices are formed in this twisted CrI$_3$ bilayer, as shown in Figs.~\ref{Fig4}(d). For the vector field of local dipoles, the AB'/BA' and AA' nodes own nontrivial winding numbers of $1$ and $-2$, respectively. Such topological invariants ensure that the morphology of ferroelectric vortices is highly robust against the structural relaxation, regardless of variations in $\theta_{\rm twist}$, thickness, and electric field (Fig.~S10 in Supplemental Material \cite{sm}). Indeed, the $2+2$ and $3+3$ twisted CrI$_3$ exhibit similar ferroelectric vortices (Fig.~S11 in Supplemental Material \cite{sm}). The only difference for these thicker cases is that the enlarged volumes reduce the polarizations which originate from the interfacial effect. However, the alternating sign of $P_z$ in AC'/CA' domains leads to zero topological charge for each ferroelectric vortices [see Fig.~\ref{Fig9}(d) in EM4], different from previously reported topological charge in twisted BN bilayer \cite{NC-2023-BN}. 

For $\theta_{\rm twist}\sim60^\circ$, the interlayer magnetic coupling is slightly different from the case of $\theta_{\rm twist}\sim0^\circ$. First, the maximum $J_L=1.72$ meV (AA') is remarkably larger than the largest one ($J_L=1.04$ meV) of $\theta_{\rm twist}\sim0^\circ$ case. As a result, the antiferromagnetic domain centers at the AA' node. Second, the sliding distance from maximum $J_L$ to $J_L=0$ is longer, leading to a larger $D=1.44$ {\AA}. Consequently, as shown in Fig.~\ref{Fig4}(e), $\theta_{\rm twist}$ within $60^\circ \pm 2.44^\circ$ can provide sufficient $d$ to accommodate antiferromagnetic bubbles in the twisted $1+1$ and $2+2$ CrI$_3$. While for the twisted $3+3$ and rigidly twisted $1+1$ CrI$_3$, the threshold angles are slightly expanded to $60^\circ \pm 2.65^\circ$. Despite the difference of $J_L$ and $D$, the threshold $d$ for antiferromagnetic bubble is almost unchanged, i.e. $\sim31$ {\AA}, since it is determined by the intralayer exchange and magnetocrystalline anisotropy.

\begin{figure}
	\centering
	\includegraphics[width=0.47\textwidth]{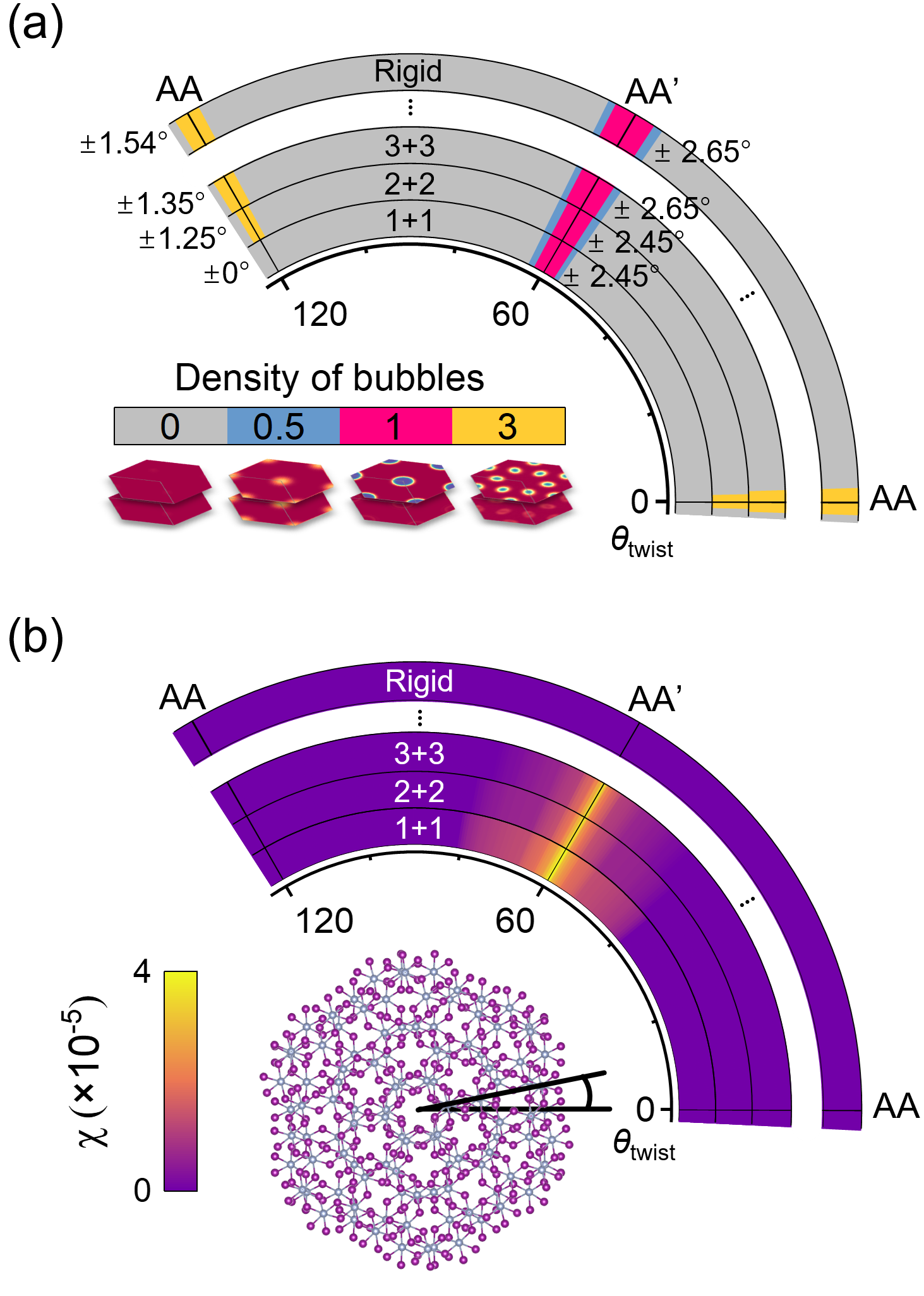}
	\caption{Summary of magnetic and ferroelectric phase diagram of twisted CrI$_3$. Evolution of (a) density of magnetic bubbles and (b) susceptibility of twisted CrI$_3$ superlattices with different $\theta_{\rm twist}$ and thickness. Although a net polarization is symmetry forbidden at zero field, an out-of-plane electric field breaks the balance of stacking domains, leading to a pronounced electric susceptibility ($\chi = \partial {P_z}/\partial {E_z}$, $E_z$: out-of-plane electric field). Larger $\chi$ appears when $\theta_{\rm twist}$ close to $60^\circ$, because of the high proportion of ferroelectric domains. }
	\label{Fig5}
\end{figure}

Another interesting issue is that this larger $J_L$ of $\theta_{\rm twist}\sim60^\circ$ can lead to a lower threshold of $d=24$ {\AA} for half antiferromagnetic bubble [Fig.~\ref{Fig4}(e)]. For a full antiferromagnetic bubble, the spins rotate for $180^\circ$ from the edge to the center of a bubble. For a half antiferromagnetic bubble, the spins rotate for $90^\circ$ from the edge to the center of a bubble. This difference between half bubble and full bubble is analogy to the relation between merons and skyrmions. As shown in Fig.~\ref{Fig4}(f), the $\theta_{\rm twist}=56.85^\circ$ $1+1$ CrI$_3$ bilayer exhibits a $d=28$ {\AA} half bubble. And the criterion of $\theta_{\rm twist}$ for half bubble also applies to the $2+2$ and $3+3$ cases due to the negligible structural relaxation effect at large $\delta$'s.

To conclude, our study has revealed that the magnetic and ferroelectric properties of twisted CrI$_3$ layers are finely controlled by the twisting angle $\theta_{\rm twist}$ and layer thickness, as summarized in Fig.~\ref{Fig5}. While the rigid lattice model can provide an initial insight, it fails to capture key results due to the vital contribution from structural relaxation, especially the suppression of magnetic bubbles in thin bilayers. In contrast, the ferroelectric vortices can survive from structural relaxation, thanks to the topological protection.Our results provide a precise theoretical description of ferroicity in twisted materials, paving the way for tailoring nontrivial ferroic orders in twisted vdW materials. Although the present study focused on CrI$_3$ only, the significance is beyond a single material. The spontaneous lattice relaxation universally exists in all twisted layers, which selectively suppresses or stabilizes different stacking regions according to their energy profiles. Then all physical properties related to stacking modes need to be understood with more realistic models beyond the rigid ones.

\begin{acknowledgments}
This work was supported by National Natural Science Foundation of China (Grants No. 12325401 and No. 12274069) and the Big Data Computing Center of Southeast University.
\end{acknowledgments}
\bibliography{apssamp}

\onecolumngrid

\vspace*{0.5cm}
\begin{center}
	{\large \bfseries End Matter}
\end{center}
\vspace*{0.5cm}

\twocolumngrid
{\it EM1. Details of structural relaxation of twisted superlattices.} Structural relaxation of the twisted superlattices is done using a lattice model, instead of a direct DFT simulation. The model total energy ($E_{\rm total}$) is consisted of the interlayer stacking energy ($E_{\rm stack}$) and intralayer elastic energy ($E_{\rm strain}$). $E_{\rm stack}$ denotes the change in energy of bilayer CrI$_3$ as a function of sliding vector $\bf u$, which can be obtained via DFT calculation, as shown in Fig.~\ref{Fig6}(a). For the sliding bilayers, only the out-of-plane coordinates of atoms are optimized in our DFT calculations, while the in-plane coordinates are fixed to avoid the change of sliding vector. As illustrated in Figs.~\ref{Fig6}(c-d), the local stacking mode in twisted superlattices is approximated to the sliding bilayers. Then, the local stacking energy [Figs.~\ref{Fig1}(c-d)] can be derived from the sliding energy landscape [i.e. Figs.~\ref{Fig6}(a-b)]. Similarly, the interlayer magnetic coupling ($J_{L}$) and local ferroelectric dipole of twisted CrI$_3$ are also approximated to the sliding cases.

\begin{figure}[H]
	\centering
	\includegraphics[width=0.47\textwidth]{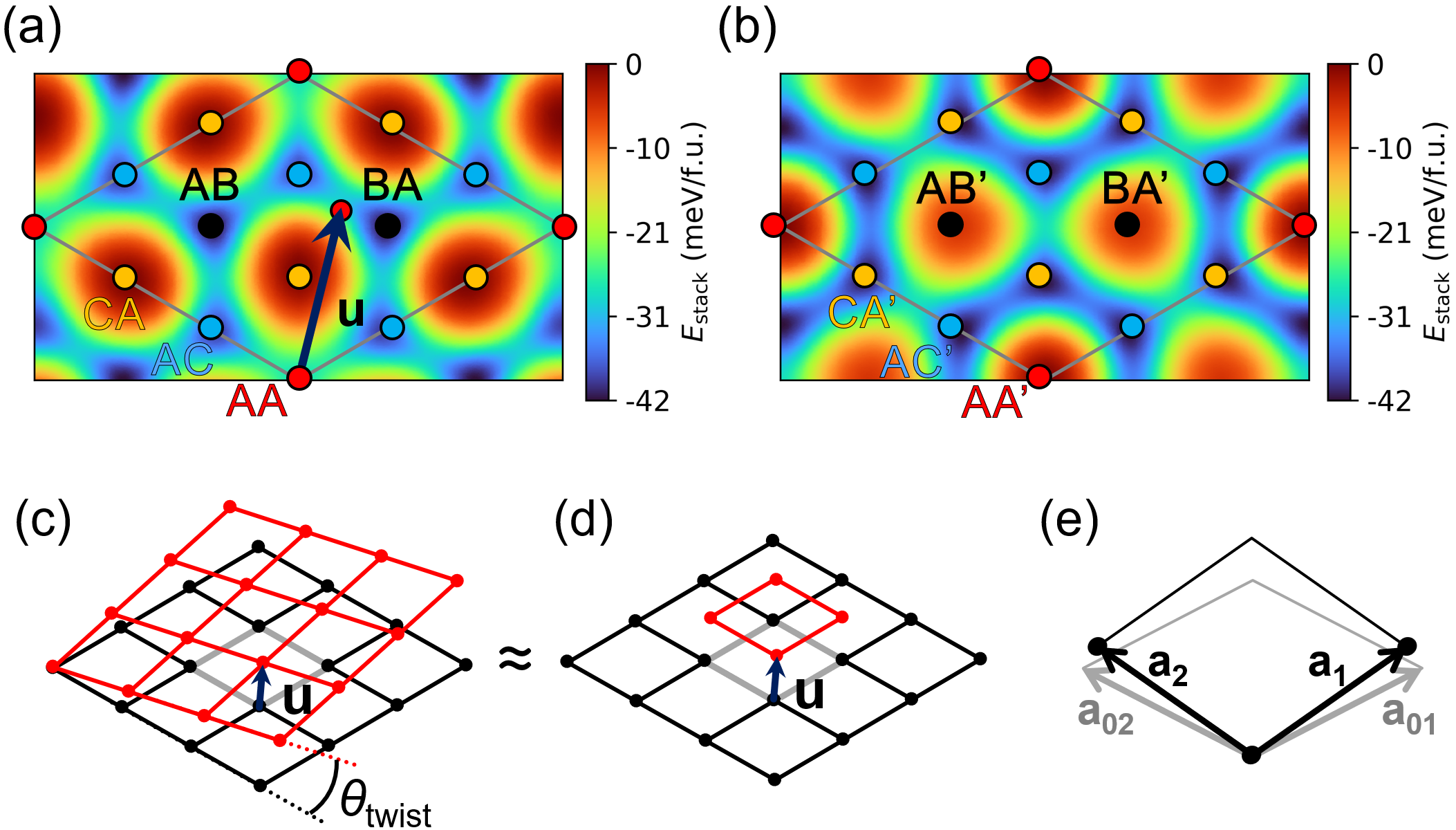}
	\caption{(a-b) Energy landscape of $E_{\rm stack}$ as a function of sliding vector $\bf u$. The original point of sliding vector $\bf u$ (i.e. ${\bf u}= 0$) is (a) the AA stacking one and (b) the AA’ stacking one. The highest energy is set as the reference $0$ for $E_{\rm stack}$. (c-d) Correspondence of sliding vector $\bf u$’s in the twisted and sliding bilayers. (e) Schematic of local lattice axes $\bf a$ (in black) and strain-free lattice axes ${\bf a}_0$ (in gray).}
	\label{Fig6}
\end{figure}

The strain tensors ($\varepsilon$) in Eq.~(\ref{E_e}) can be obtained from \cite{b-strain}:
\begin{equation}
	\varepsilon  = {\bf a}{\bf a}_0^- - \bf{I} = \left[ {\begin{array}{*{20}{c}}
			{{\varepsilon _{11}}}&{{\varepsilon _{12}}}\\
			{{\varepsilon _{21}}}&{{\varepsilon _{22}}}
	\end{array}} \right],
\end{equation}

where $\bf a$ is the local lattice axes and ${\bf a}_0$ is the strain-free lattice axes as shown in Fig.~\ref{Fig6}(e). $\bf I$ is the identity matrix. The strain tensor is reshaped to $\varepsilon = [\varepsilon_{11}, \varepsilon_{22}, \varepsilon_{12} + \varepsilon_{21}]$ to match the elastic tensors:

\begin{equation}
	C = \left[ {\begin{array}{*{20}{c}}
			{{C_{11}}}&{{C_{12}}}&0\\
			{{C_{21}}}&{{C_{11}}}&0\\
			0&0&{{C_{66}}}
	\end{array}} \right],
\end{equation}
where $C_{66} = (C_{11} - C_{12})/2$. The elastic tensors are obtained by fitting the energy-strain curves to parabolic equation of $E_{\rm strain}$ as shown in Fig.~S2(a) in Supplemental Material and the results are listed in Table.~S1 in Supplemental Material \cite{sm}.

{\it EM2. Details of micromagnetic simulation.} In the micromagnetic simulation, the spin dynamics are described by the Landau-Lifshitz-Gilbert (LLG) equation:
\begin{equation}
	\frac{{\partial {{\bf S}_i}}}{{\partial t}} = -\frac{\gamma }{{(1 + {\alpha ^2})}}{\bf S}_i \times ({\bf H}_{\rm eff}^i + \alpha {\bf S}_i \times {\bf H}_{\rm eff}^i),
\end{equation}
where $\alpha$ is the Gilbert damping constant, $\gamma$ is the Gilbert gyromagnetic ratio, and ${\bf H}_{\rm eff}^i=- \frac{1}{{{\mu_s}}}\frac{{\partial H}}{{\partial {\bf S}_i}}$ is the effective field. $\mu_S = 3 ~ \mu_B$ is the magnetic moment of Cr. The LLG equation is solved by the fourth order Runge-Kutta method. ${\bf H}_{\rm eff}^i$ is derived from the Hamiltonian:
\begin{equation}
	\begin{aligned}
		H &= J_{ij}\sum_{\langle i,j\rangle} {{{\bf{S}}_i} \cdot {{\bf{S}}_j}}  - K\sum\limits_i {{{({S}_i^z)}^2}} \\
		&- \frac {\mu_0 \mu_s^2} {4\pi} \sum\limits_{i \neq j}{\frac{{3({{\bf{S}}_i} \cdot {{\bf{r}}_{ij}})({{\bf{S}}_j} \cdot {{\bf{r}}_{ij}}) - {{\bf{S}}_i} \cdot {{\bf{S}}_j}}}{{{{\bf{r}}_{ij}}}}},
		\label{ham}
	\end{aligned}
\end{equation}
with $J$ and $K$ as the magnetic coupling parameters and the magnetocrystalline anisotropy energy, and the third term is the demagnetization energy.

\begin{figure}[H]
	\centering
	\includegraphics[width=0.47\textwidth]{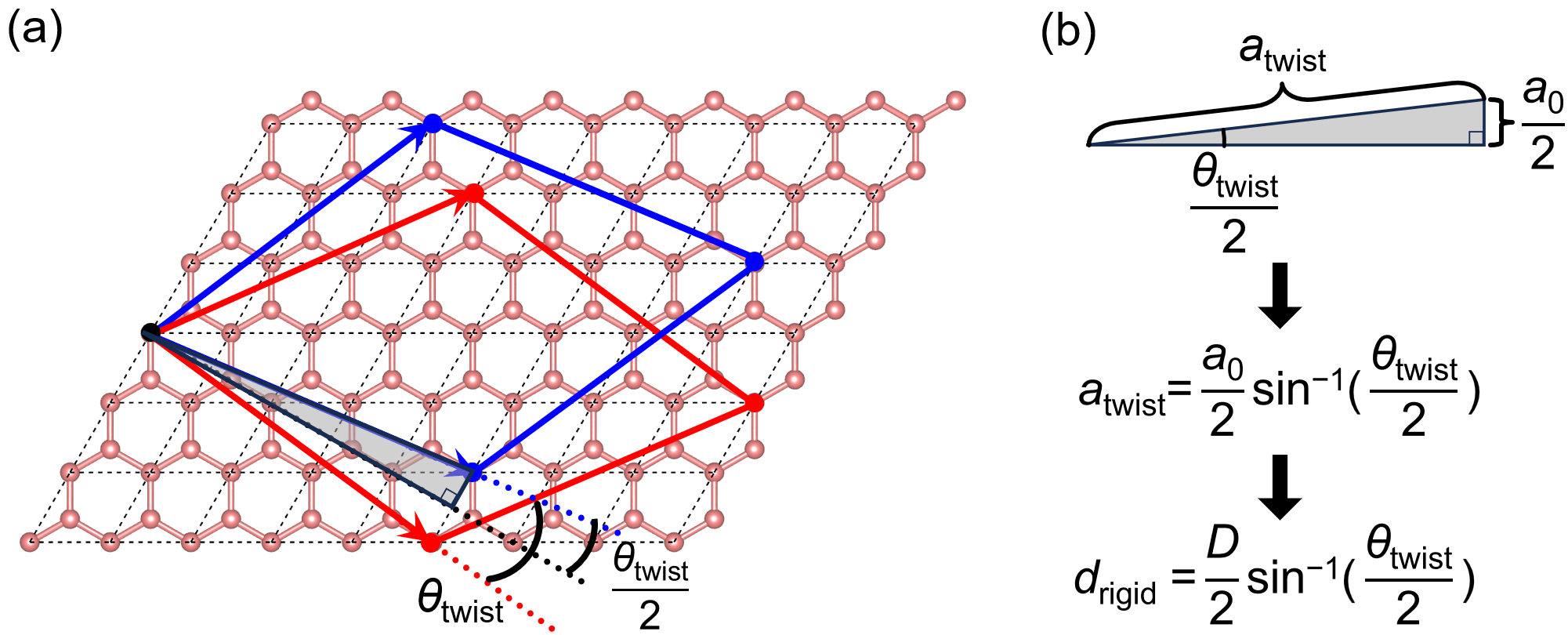}
	\caption{(a) Schematic of constructing a twisted superlattice. (b) Geometric triangulation relationship between the twisted superlattice and unit cell.}
	\label{Fig7}
\end{figure}

\begin{figure}[H]
	\centering
	\includegraphics[width=0.47\textwidth]{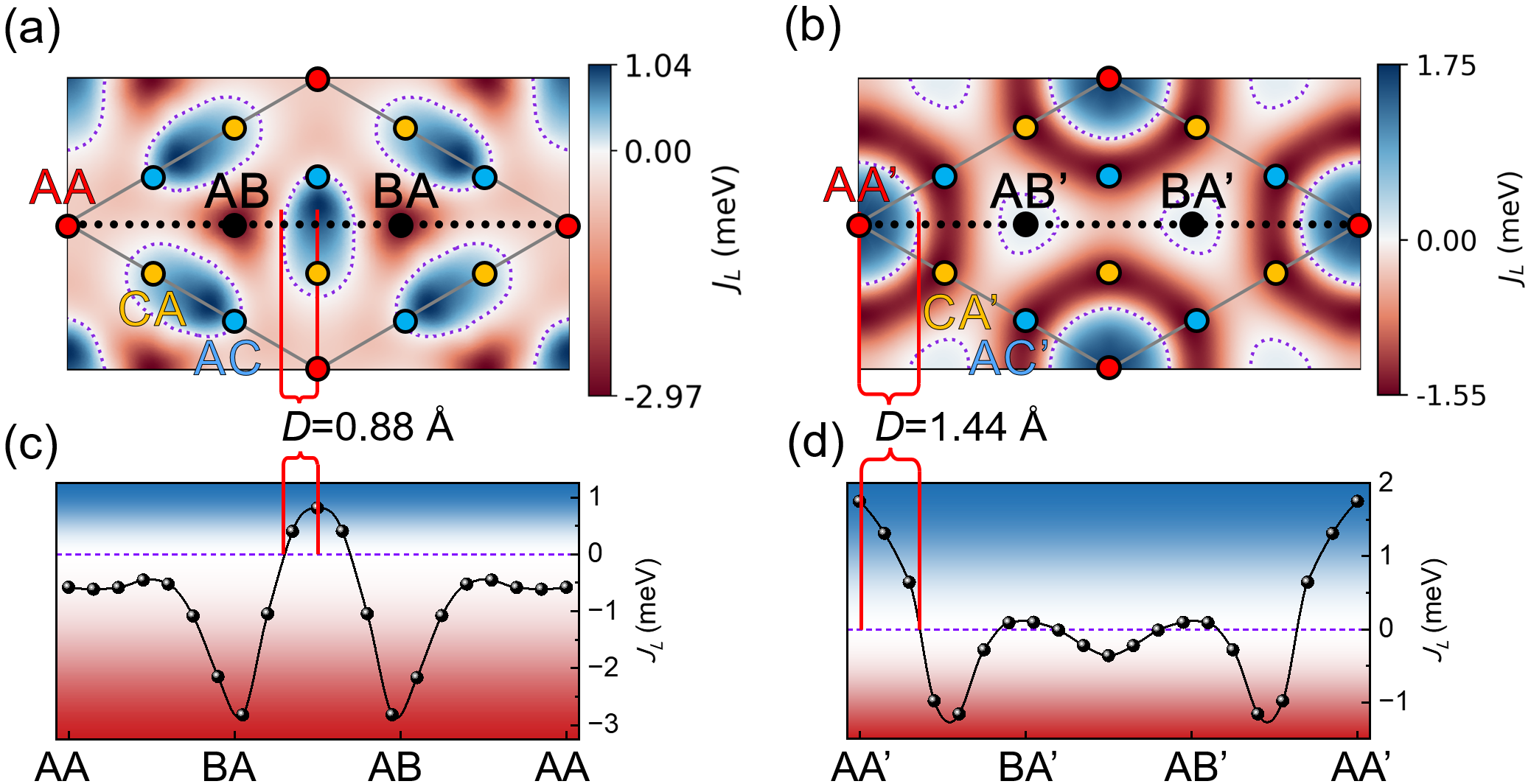}
	\caption{(a-b) The interlayer magnetic coupling ($J_L$) of sliding CrI$_3$ bilayers as a function of sliding vector $\bf u$. (a) The AA stacking and its sliding derivatives. (b) The AA' stacking and its sliding derivatives. (c-d) The evolution of $J_L$'s along the black dash lines in (a-b). $D$ is defined as the half-width of sliding distance corresponding to the antiferromagnetic coupling. Purple broken curve: the $J_L=0$ boundary.}
	\label{Fig8}
\end{figure}

\begin{figure}[H]
	\centering
	\includegraphics[width=0.47\textwidth]{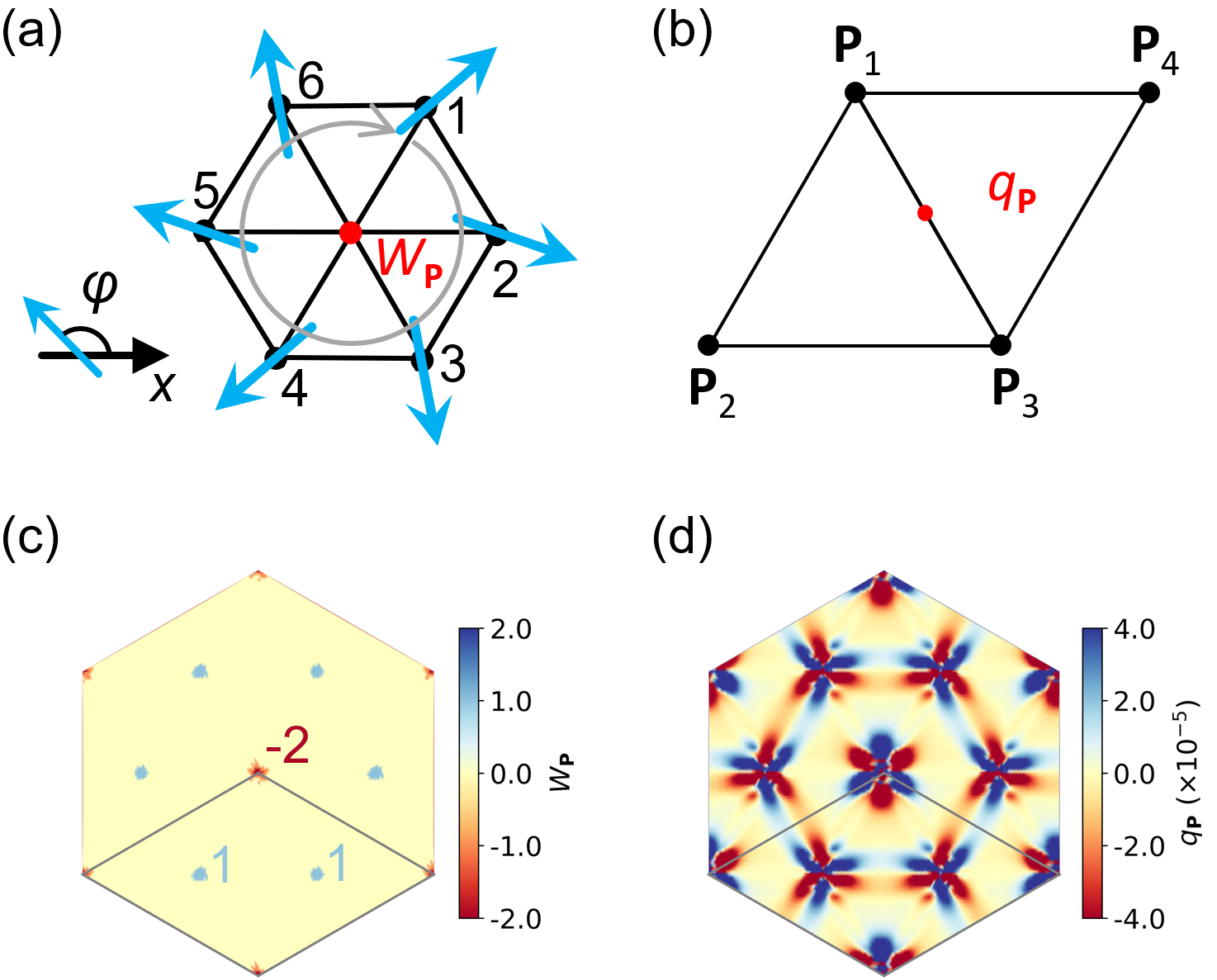}
	\caption{Definition of (a) azimuth angles $\varphi$ and (b) topological charge $q_{\bf P}$. Contour maps of (c) winding number $W_{\bf P}$ and (d) topological charge $q_{\bf P}$ of $59.01^\circ$ $1+1$ CrI$_3$ with structural relaxation.}
	\label{Fig9}
\end{figure}

{\it EM3. Geometric triangulation in twisted superlattices.} Eq.~(\ref{d}) stems from the relationship between lattice constants of twisted superlattice ($a_{\rm twist}$) and unit cell ($a_0$). As shown in Fig.~\ref{Fig7}(a), a $\theta_{\rm twist}$ twisted superlattice is constructed by two supercells with $\pm\theta_{\rm twist}/2$. Then, $a_{\rm twist}$ can be obtained according to the geometric triangulation relationship as shown in Fig.~\ref{Fig7}(b). If $a_0$ is replaced by the sliding distance ($D$) between two stacking modes in sliding bilayers (see Fig.~\ref{Fig8}), the spatial distance ($d_{\rm rigid}$) between these two stacking modes in the rigidly twisted superlattices is obtained.

{\it EM4. Details of topological properties calculations.} The winding number ($W_{\bf P}$) is calculated by the change in azimuth angles ($\varphi$) of dipoles around a local site as shown in Fig.~\ref{Fig9}(a):
\begin{equation}
\begin{aligned}
	W_{\mathbf P}
	= \frac{1}{2\pi}
	\sum_{i=1}^{6} \bigl(\varphi_{i+1} - \varphi_i\bigr)
	\,,\qquad
	\varphi_{7} = \varphi_1 \,.
\end{aligned}
\end{equation}

The local topological charge ($q_{\bf P}$) of polarization is calculated via the solid angles form by four nearest unit cells \cite{NPB-1981-q}:
\begin{equation}
\begin{aligned}
	q_{\bf{P}} = \frac{1}{2\pi} \bigl[
	\arg[1 + \bf{P}_1 \!\cdot\! \bf{P}_2 + \bf{P}_2 \!\cdot\! \bf{P}_3 + \bf{P}_3 \!\cdot\! \bf{P}_1 \\
	+ i \bf{P}_1 \!\times\! (\bf{P}_2 \!\times\! \bf{P}_3)] \\
	+ \arg[1 + \bf{P}_1 \!\cdot\! \bf{P}_3 + \bf{P}_3 \!\cdot\! \bf{P}_4 + \bf{P}_4 \!\cdot\! \bf{P}_1 \\
	+ i \bf{P}_1 \!\times\! (\bf{P}_3 \!\times\! \bf{P}_4)] \bigr]
\end{aligned}
\end{equation}
, where ${\bf P}_i$ is the normalized polarization of a local unit cell and a solid angle of a local site is shown in Fig.~\ref{Fig9} (b).

\end{document}